%Paper: hep-th/9407192
%From: klemm@nxth04.cern.ch
%Date: Fri, 29 Jul 94 11:43:34 +0100

\input harvmac
\def\npb#1(#2)#3{{ Nucl. Phys. }{B#1} (#2) #3}
\def\plb#1(#2)#3{{ Phys. Lett. }{#1B} (#2) #3}
\def\pla#1(#2)#3{{ Phys. Lett. }{#1A} (#2) #3}
\def\prl#1(#2)#3{{ Phys. Rev. Lett. }{#1} (#2) #3}
\def\mpla#1(#2)#3{{ Mod. Phys. Lett. }{A#1} (#2) #3}
\def\ijmpa#1(#2)#3{{ Int. J. Mod. Phys. }{A#1} (#2) #3}
\def\cmp#1(#2)#3{{ Commun. Math. Phys. }{#1} (#2) #3}
\def\cqg#1(#2)#3{{ Class. Quantum Grav. }{#1} (#2) #3}
\def\jmp#1(#2)#3{{ J. Math. Phys. }{#1} (#2) #3}
\def\anp#1(#2)#3{{ Ann. Phys. }{#1} (#2) #3}
\def\prd#1(#2)#3{{ Phys. Rev.} {D\bf{#1}} (#2) #3}

\def\p{\partial}

\def\da#1{{\p \over \p s_{#1}}}
\def\inbar{\,\vrule height1.5ex width.4pt depth0pt}
\def\IQ{\relax\,\hbox{$\inbar\kern-.3em{\rm Q}$}}
\def\IB{\relax{\rm I\kern-.18em B}}
\def\IC{\relax\hbox{$\inbar\kern-.3em{\rm C}$}}
\def\IP{\relax{\rm I\kern-.18em P}}
\def\IR{\relax{\rm I\kern-.18em R}}
\def\ZZ{\relax\ifmmode\mathchoice
{\hbox{Z\kern-.4em Z}}{\hbox{Z\kern-.4em Z}}
{\lower.9pt\hbox{Z\kern-.4em Z}}
{\lower1.2pt\hbox{Z\kern-.4em Z}}\else{Z\kern-.4em Z}\fi}
\def\hb{\hfill\break}

% when printing final draft, remove \draftmode
%\draftmode
\rightline{CERN-TH.7369/94}
\vskip -14mm

\Title{}{A Note on ODEs from Mirror Symmetry
\footnote{${}^\diamond$}{Research supported by grant DE-FG02-88-ER-25065.}}

%For more complicated situations, substitute for {\it either\/} argument:
%\Title{\vbox{\baselineskip12pt
%              \hbox{Not for distribution}}}
%{\vbox{\centerline{Untitled}}
%       \vskip2pt\centerline{}}}
%   \footnote{}{}
\centerline{
A. Klemm$^1$,
B.H. Lian$^2$,
S.S Roan$^3$
and S.T Yau$^4$\footnote{ }{email:  klemm@nxth21.cern.ch,
lian@math.harvard.edu, maroan@ccvax.sinica.edu.tw and \hb
                                                  \phantom{XX email : }
yau@math.harvard.edu}}

\bigskip
\centerline{$^1$
            \it Theory Divison,}
\centerline{\it CERN, CH-1211 Geneva 23}
\bigskip
\centerline{$^{2,4}$
            \it Department of Mathematics}
\centerline{\it Harvard University}
\centerline{\it Cambridge, MA 02138, USA}
\bigskip
\centerline{$^3$
            \it Institute of Mathematics}
\centerline{\it Academia Sinica}
\centerline{\it Taipei , Taiwan}
\vskip .5in
\centerline{\it In Honor of Professor Israel M. Gel$'$fand on the occasion of
his $80^{th}$ birthday.}

\vskip .3in

Abstract: We give close formulas for the counting functions of rational
curves on complete intersection Calabi-Yau manifolds in terms of special
solutions of generalized hypergeometric differential systems. For the
one modulus cases we derive a differential equation for the Mirror map,
which can be viewed as a generalization of the Schwarzian equation.
We also derive a nonlinear seventh order differential equation
which directly governs the instanton corrected Yukawa coupling.
\Date{CERN -TH. 7369, July 1994}

\newsec{Introduction}
In a seminal paper \ref\cdgp{P. Candelas, X. della Ossa, P. Green,
L. Parkes, \npb359(1991)21} physicists solved a problem in
enumerative geometry, namely to count\foot{Only if
the moduli space of the map from $\IP^1$
to $X$, with three points fixed,
is zero dimensional $n_d$ counts the number of isolated
rational curves. In general $n_d$ has to be understood as the integral
of the top Chern class of a vectorbundle over the moduli space of that map.}
the ``number'' $n_d$ of
rational curves\foot{These are called instantons by physicists as
they correspond to classical solutions of the $\sigma$-model
equations of motion.} of arbitrary degree $d$ on the quintic threefold $X$ in
$\IP^4$. The answer was given in terms of the large volume
expansion of the correlation function, called Yukawa coupling by physicists,
between three states ${\cal O}_{J}$  of the twisted $N=2$
topological $\sigma$-model on $Q$, which has the formal
expansion
\cdgp\ref\am{P. Aspinwall and D. Morrison,\cmp151(1993) 45}\ref\witten{E.
Witten,
\cmp118(1988)411;
\npb371(1992)191 and in Essays on Mirror Manifolds, Ed. S.-T. Yau,
(International Press, Hong Kong 1992)}
\eqn\yukawaI{
K_{JJJ}=
\langle {\cal O}_{J}{\cal O}_{J}{\cal O}_{J}\rangle
=\int_X J\wedge J\wedge J + \sum_{d} {d^3 n_d q^d \over(1-q^d)}.}
Here $q=e^{2 \pi i t}$ and the modulus $t$ parametrizes
the complexified K\"ahler class of $X$,
i.e. $({\rm Im} (t))^3\propto$ volume of $X$ and ${\rm Re}(t)$
parametrizes an antisymmetric tensor field, which is the component of a
harmonic $(1,1)$-form on $X$
\ref\rw{R. Rohm and E. Witten, \anp234(1987)454}.
The correlation function \yukawaI, is sometimes
referred to as an intersection  of the quantum (co)homology of $X$.
In the large volume limit the contribution of the
instantons is damped out and \yukawaI~approaches the classical
self intersection number between the cycle dual to the K\"ahler form $J$.

It is a remarkable fact, that this counting function \yukawaI~is
expressible in a closed form in terms of solutions of
a generalized hypergeometric system. This has been used in
\ref\bs{V. Batyrev and D. van Straten, {\sl Generalized
hypergeometric functions and rational curves on Calabi-Yau
complete intersections in toric varieties},
Essen Preprint (1993)}\ref\hktyI{S. Hosono, A. Klemm, S. Theisen and S.-T. Yau:
           {\sl Mirror symmetry, mirror map and applications
           to Calabi-Yau hypersurfaces},
           HUTMP-93/0801, LMU-TPW-93-22  (hep-th/9308122), to be
           published in Commun. Math. Phys.}\ref\hktyII{S. Hosono, A. Klemm,
           S.Theisen and S. T. Yau, {\sl Mirror symmetry, mirror map
           and applications to complete intersection Calabi-Yau
           spaces\/}, Preprint HUTMP-94-02, hep-th 9406055}
to predict the number of rational curves on various Calabi-Yau spaces.

In the section 2 of this exposition we review the physical reasoning,
which explains that fact and give, as a generalization,
closed formulas for the counting functions on nonsingular complete
intersection Calabi-Yau spaces in products of weighted projective spaces.
An important step in these calculations is the definition of the mirror
map. We discuss therefore in section 3 and 4 the differential
equation which governs the mirror map. As we will see in section 2
the most important quantity is the prepotential, from which the
the correlation function \yukawaI~ and the Weil-Peterson metric for the complex
moduli space,
can be derived. We will obtain in section 5 a differential
equation for the prepotential.

\newsec{Counting of rational curves and generalized hypergeometric functions}
It was argued by Witten \witten~that the states in the ($N=2$)
topological $\sigma$-model on an (arbitrary Calabi-Yau) space $X$ are in
one to one correspondence with the elements in the cohomology groups of
$X$, for example the
state ${\cal O}_J$ above correspond to
the K\"ahler form in $H^{1,1}(X,{\bf Z})$.
As there is a natural involution symmetry in the $N=2$ topological
$\sigma$-model on Calabi-Yau manifolds
exchanging the states corresponding to the
cohomology groups $H^{3-p,q}(X)$ and $H^{p,q}(X)$,
physicists suspect that Calabi-Yau spaces occur quite
generally in mirror pairs\foot{See \ref\batyrev{V. Batyrev,
{\sl Dual Polyhedra and the Mirror Symmetry
      for Calabi-Yau Hypersurfaces in Toric Varieties}, Univ. Essen
      Preprint (1992), to appear in  Journal of Alg. Geom.}\hktyI~and
references  therein for geometrical constructions of mirror pairs,
which support this expectation.}
$X$ and $X^*$, in which the r\^ole of these cohomology groups are
exchanged. In particular $h^{3-p,q}(X)=h^{p,q}(X^*)$ holds and the
Euler number of $X$ is therefore the negative of the
Euler number of $X^*$.
By the same token it is expected that the correlation functions among one
type of states on $X$ can be calculated by the same methods
as its counterparts among the corresponding states on $X^*$.
In particular the correlation function \yukawaI~is related
by this argument\cdgp\witten~to the correlation function
\eqn\yukawaII{
K_{J^*J^*J^*}=
\langle {\cal O}_{J^*} {\cal O}_{J^*} {\cal O}_{J^*}\rangle=
\int_{X^*} \Omega\wedge b^\alpha_{J^*}\wedge b^\beta_{J^*}\wedge b^\gamma_{J^*}
\Omega_{\alpha\beta\gamma},}
where $\Omega$ is the unique no-where vanishing holomorphic
threeform on $X^*$ and $b^\alpha_{J^*}\in H^1(X,TX)\cong H^{2,1}(X)$.
In fact the expansion \yukawaI~and hence the successful prediction
of rational curves on the quintic $X$ was obtained in
\cdgp~by calculating the correlation function \yukawaII~on the
mirror $X^*$ using classical methods of the theory of
complex structure deformation.
The integral on the right-hand side of \yukawaII, introduced in
\ref\bg{R. Bryant and P. Griffiths, Progress in Mathematics {\bf 36} 77
(Birkh\"auser Boston 1983)} depends only, via $\Omega$, on the
choice of the complex structure modulus on $X^*$  but not on the choice of
complexified K\"ahler moduli on $X^*$, while \yukawaI~depends only on the
complexified K\"ahler
modulus on $X$, but not on the complex  structure moduli of $X$. Due to mirror
symmetry we should be able to identify
the complex structure modulus space of $X^*$ with the complexified K\"ahler
structure  modulus space of $X$.
Hence after calculating the dependence of \yukawaII~on the complex structure
modulus of $X^*$
using the Picard-Fuchs equation, the decisive steps are to find the point of
expansion in
this modulus space of  $X^*$, which corresponds to the large volume limit of
$X$,
and to determine the  map from the complex  structure modulus of $X^*$ in
\yukawaII~to the K\"ahler
structure modulus $t$ of $X$ in \yukawaI. This map (or its inverse) will be
called the mirror map.

For the quintic hypersurface in ${\IP}^4$, the mirror $X^*$
can be constructed concretely as the canonical desingularized quotient
$X^*={\widehat {X/({\ZZ }_5^3)}}$, where ${\ZZ}_5^3$ acts by
phase multiplication on the homogenoeus coordinates
$(x_1:\ldots:x_5)$ of the ${\IP}^4$ and is generated by
$g_i:(x_i,x_5) \mapsto (\exp(2 \pi i/5)x_i,\exp(8 \pi i/5)x_5)$
$i=1,2,3$.
Here \yukawaII~depends on the one-dimensional complex structure
deformation of $X^*$, which can be studied
by considering the deformations of the quintic $X$,
but restricted to the unique ${\ZZ}_5^3$-invariant
element $x_1x_2x_3x_4x_5$ in its local ring.

The Picard-Fuchs ODE can therefore be derived by the
Dwork-Griffith-Katz reduction method from the standard residuum
expression of the period \ref\Griffiths{P. Griffiths, Periods of integrals on
algebraic manifolds, I and II., Amer. Jour. Math. vol. 90 (1968), 568., Ann.
Math. 90 (1969) 460.} for $N=5$
\eqn\period{\tilde\omega_i(s_0,\ldots,s_N)=\int_\gamma\int_{\Gamma_i}
{\mu\over \sum_{i=1}^N s_i x_i^N - s_0 \prod_{i=1}^N x_i},}
where $\gamma$ is a small cycle in ${\IP}^{N-1}$, $\Gamma_i\in H_{N-2}(M)$
and  the measure is
$$\mu=\sum_i (-1)^i x_i dx_1\wedge \ldots
{\widehat {dx_i}}\ldots \wedge dx_N.$$
Instead of using this generic
alogarithm, let us consider of the symmetries of \period~directly.
Obviously
\eqnn\globsyma
\eqnn\globsymb
$$
\eqalignno{
& \hat\omega_i(\lambda^N s_0,\ldots,\lambda^N s_N)=
\lambda^{-N}\hat\omega_i(s_0,\ldots,s_N)
& \globsyma \cr
&\hat\omega_i(s_0,\ldots,\lambda_i^N,\ldots,\lambda_i^{-N} s_N)=
\hat\omega_i(s_0,\ldots,s_N),\quad {\rm for}\;\; i=1,\ldots,N-1,
& \globsymb \cr
}
$$
with $\lambda,\lambda_i\in {\bf C}^*$. Writing \globsyma, \globsymb~
in infinetessimal form we obtain the differential equations
\eqnn\infsyma
\eqnn\infsymb
$$
\eqalignno{
& \left\{ \sum_{i=0}^{N}s_i\da{i}+1 \right\} \hat\omega_i(s)=0  \quad ,
& \infsyma \cr
& \left( s_i\da{i}-s_N\da{N} \right)\hat\omega_i(s) =0
 \;\; {\rm for}\;\; i=1,\cdots ,N-1 \;.
& \infsymb \cr
}
$$
The trivial relation $x_1^N\ldots x_N^N-(x_1\ldots z_N)^N\equiv 0$ leads
to a further differential equation
\eqn\Nsym{
\left\{
\prod_{i=1}^N\da{i} - \left(\da{0}\right)^N\right\}\hat\omega_{i}(s)=0 \, .
}
This system of differential equation \infsyma-\Nsym~
is precisely the type of generalized hypergeometric system,
which was investigated by Gel'fand, Kapranov and Zelevinskii in
\ref\gkzI{I. M. Gel'fand, A. V. Zelevinkii and M. M. Kapranov,
Hypergeometric functions and toral manifolds, Functional Anal. Appl.
{\bf 23} 2 (1989) 12, English trans. 94} with the characters defined by
\eqn\chis{
\eqalign{
\chi_1&=(1,\underbrace{0,\ldots,0}_{(N-1)-times}),
\quad\chi_2=(1,1,0,\ldots,0),\ldots\cr
&\ldots, \chi_N=(1,0,\ldots,0,1),\quad
\chi_{N+1}=(1,\underbrace{-1,\ldots,-1}_{(N-1)-times})}}
and the exponents: $\vec \beta=(-1,0,\ldots,0)$.
The generator of the lattice ${\bf L}$ of relations
is \eqn\generator{l=(-N;\underbrace{1,\ldots,1}_{N-times}).}
The eqns. \infsyma,\infsymb~are satisfied identically by the ansatz
\eqn\ansatz{\hat \omega_i(s)={1\over s_0} \omega
\left(\prod_{i=1}^N s_i\over s_0^N\right).}
By using the new coordinate for the
complex structure modulus
\eqn\newvar{z=(-1)^{l_0} \prod_{i=0}^N s_i^{l_i}}
the eqn. \Nsym~can be brought in the following convenient form
\eqn\pfquintic{\theta_z\left[\theta^{N-1}-N z\,
\prod_{i=1}^{N-1} (N \theta+i)\right]\,\omega_i=0,} where $\theta=z {d\over
dz}$. The generalized hypergeometric system defined by \chis~and
$\vec \beta$ is proven to be holonomic
\gkzI~and a formal powerseries expansion and
(Euler) integral representations were likewise given.
For the quintic ($N=5$) the system has 5 solutions, but it
is semi-resonant, which implies that the monodromy on the full solutions
space is reducible. On the other hand the monodromy for the 4 periods on $Q^*$
is known to be irreducible. The unique subsystem of the solutions
of \pfquintic~on which the monodromy acts irreducible
is given by the 4 solutions to
\eqn\pfN{\left[\theta^{N-1}-N z\,
\prod_{i=1}^{N-1} (N \theta+i)\right]\,\omega_i=0,}
which identifies the later equation with the Picard-Fuchs
equation of the mirror $X^*$.

The complex structure moduli space of a Calabi-Yau threefold exhibits
special geometry, as it was explained in
\ref\specialgeom{A. Strominger, \cmp133(1990)163;  P. Candelas and X. della
Ossa,\npb355(1991)455} using crucially the results
of \ref\tian{G. Tian, in {\sl Mathematical Aspects of String
Theory}, ed. S. T. Yau, (World Scientific, Singapore 1987)}.
This structure is charaterized by the existence of
a section $\tilde F$ of a holomorphic line bundle over the
complex moduli space, which is a prepotential for structure contant(s)
\yukawaII~and the K\"ahler potential $K$ of the Weil-Peterson metric.
There exists a special coordinate choice, given by a ratio of periods
$\tilde t=\tilde \omega_1(z)/\tilde \omega_0(z)$ in which these relations read
\eqn\derived{\eqalign{
K_{J^*J^*J^*}&=\partial^3_{\tilde t} {\tilde F} \cr
K&=-\log \left((\tilde t-\bar {\tilde t}) (\partial_{\tilde t} {\tilde F}+
\bar\partial_{\tilde t} \bar {\tilde F})+
2\,\bar {\tilde F}- 2\,{\tilde F} \right).}}
These coordinates can equivalently be characterized by the property that
the period vector is expressible in terms of the prepotential as
\eqn\periods{\Pi(z) =(\tilde \omega_0(z),\tilde \omega_1(z),
\tilde \omega_2(z),\tilde \omega_3(z))=\tilde \omega_0
(1,\tilde t,\partial_{\tilde t} {\tilde F},2{\tilde F}-
\tilde t \partial_{\tilde t} {\tilde F})}
and vice versa
\eqn\dumb{\tilde F(\tilde t)={1\over 2 \tilde\omega_0^2}(\tilde\omega_3
\tilde\omega_0+
\tilde\omega_1\tilde\omega_2).}

It has been argued \cdgp\specialgeom that the moduli space
of the complexified K\"ahler structure of the the $N=2$ topological
$\sigma$-model exhibits also special geometry with \yukawaI~as
structure constant(s) and that $t$ is the
special coordinate (especially for the last point see also
\ref\bcovII{M. Bershadsky, S. Ceccotti, H. Ooguri and
C. Vafa {\sl Kodaira-Spencer theory of gravity and exact results
for quantum string amplitudes}, HUTP-93/A025, RIMS-
946, SISSA-142/93/EP}).
Because of the analog of \derived~for the K\"ahler structure modulus
the prepotential $F(t)$ is determined by $K_{JJJ}$ up to a quadratic
polynomial in $t$:
\eqn\prepot{F(t)= {\int_X J\wedge J\wedge J \over 3!}t^3+ {a\over 2} t^2+b t+c
+ F_{inst}(q).}
To identify $t$ with $\tilde t$ and $F(t)$ with $\tilde F(\tilde t)$,
we must find in the complex structure moduli
space of $X^*$ the special point $z_1$, which corresponds
to the large volume limit ${\rm Im}(t)\rightarrow \infty$ of $X$.
This can be done in the following
heuristic way. First note the invariance of \yukawaI~under the
shift symmetry $t\rightarrow t+1$. In fact more generally,
shifting the parameter
of the antisymmetric background ${\rm Re}(t)$ by an integer is a symmetry
of the $\sigma$-model in the large volume region\witten\rw.
We require that the transformation of the ``period''
$(1,t,\p_t F, 2 F-t \p_t F)$ under that symmetry should
correspond to a monodromy operation on $\Pi(z)$ under counterclockwise
analytic continuation around $z_1$.
That is, we search a point $z=z_1$ in the complex modulus space
with the specific monodromy action:
\eqn\monodromy{{\vec \Pi}(z)\mapsto
\left(\matrix{            1&                         0&\phantom{-}0&0\cr
                         1&                         1&\phantom{-}0&0\cr
 \left(a+{K\over 2}\right)&                         K&\phantom{-}1&0\cr
 \left(2b-{K\over 6}\right)&\left(a-{K\over 2}\right)&        -1&1}\right)
{\vec \Pi} (z),\quad {\rm with}\,\, K=\int_X J\wedge J\wedge J}
which is unipotent of order 4. The importance of this monodromy
requirement was pointed out in \ref\morrison{D. R. Morrison,  in Essays
on mirror manifolds Ed. S.-T.Yau (International Press Singapore 1992)}.
It is easy to see that among the three regular singular points
$z=0,1/{5^5},\infty$ of the ODE \pfN~with $N=5$, the point that admits
such a monodromy is $z=0$,
where the indicial equation is four-fold degenerate.
Around this point, there is one powerseries solution given by
\eqn\powsol{\omega_0(z)= \omega_0(z,\rho)|_{\rho=0} =
\left. { \sum_{n\ge 0} c(n,\rho)z^{n+\rho} }\right|_{\rho=0},}
where the coefficients $c(\rho,n)$ can be expressed in terms of
gamma-functions from the $l$ in \generator~as
\eqn\coefficients{c(n,\rho)={\Gamma\left(l_0 (n+\rho)+1\right)
                             \over \prod_{i=1}^N
                             \Gamma\left(l_i(n+\rho)+1)\right)}.}
The other solutions can be obtained by the well-known Frobenius
method (see e.g. \ref\Yoshida{M. Yoshida, Fuchsian Differential Equations,
Frier. Vieweg \& Sohn, Bonn 1987.}):
\eqn\logsolutions{\omega_p=\left. { {1\over p !}
\left({1\over 2 \pi i} {\partial\over {\partial_\rho} }\right)^p
\omega_0(z,\rho) }\right|_{\rho=0},\quad {\rm for } \,\,\, p=1,\ldots,N-2\, .}
Their monodromy is dictated by the terms linear, quadratic
und cubic in $\log (z)$. By comparing the monodromy of
these solutions with \monodromy~we
conclude that the mirror map is given by
\eqn\mirrormap{t={\omega_1(z)\over \omega_0(z)}.}
Also from the monodromy requirement and using the special geometry
relations \derived~we get, indepentently of $a,b,c$, a unique
expansion of \yukawaI~completely expressed in terms of
special solutions to the GKZ system :
\eqn\yukawaexp{K_{JJJ}={\int_X J\wedge J\wedge J \over 2}
                 \partial^2_t\left(\omega_2(z(t))\over \omega_0(z(t))\right),}
where we denote (the inverse of) the mirror map \mirrormap~by $z(t)$.

Let us finish this section with the generalization of the result
\yukawaexp~to nonsingular complete intersection Calabi-Yau spaces
in products of $k$ weighted projective spaces and give closed formulas
for the large radius expansions of the triple intersection \yukawaI~$\langle
{\cal O}_{J_i} {\cal O}_{J_j} {\cal O}_{J_k}\rangle $, where $J_i$ is the
K\"ahler class induced from the $i$th weighted projective space.
{}From these expansions one can read off the numbers of rational
curves of any multidegree spaces, with respect to the K\"ahler classes
induced from the projective spaces. These results were obtained in \hktyII.

We consider in the following complete intersections of $l$
hypersurfaces in products of $k$  projective spaces.
Since most formulas allow for an incorporation of
weights we will state them for the general case.
Denote by $d_j^{(i)}$ the degree of the coordinates of
$\IP^{n_i}[\vec w^{(i)}]$ in the $j$-th polynomial $p_j$
($i=1,\dots,k;\,\,j=1,\dots,l$).
The residuum expression for the periods \Griffiths, with
$k$ perturbations satisfies again a GKZ-system, where the
lattice of relations ${\bf L}$
is generated by $k$ generators $l^{(s)}$ $(s=1,\ldots,k,j=1,\ldots,l)$
\eqn\cpislor{l^{(s)}=(-d_1^{(s)},\ldots,-d_l^{(s)};
\ldots,w_1^{(s)},\ldots,w_{n_s+1},0,\ldots)\equiv
\left(\{l^{(s)}_{0,j}\};\{l_i^{(s)}\}\right),}
from which one obtains $k$ linear differential operators
($\theta_s=z_s {d\over d z_s}$)
\eqn\lindiffop{\eqalign{
{\cal L}_s=&\prod_{j=1}^{n_s+1}\Bigl(w_j^{(s)}
\theta_s\Bigr)
\Bigl(w_j^{(s)}\theta_s-1\Bigr)\cdots
\Bigl(w_j^{(s)}\theta_s-w_j^{(s)}+1\Bigr) \cr
&\qquad -\prod_{j=1}^l\Bigl(\sum_{i=1}^k d_j^{(i)}\theta_i\Bigr)
\cdots\Bigl(\sum_{i=1}^k d_j^{(i)}\theta_i-d_j^{(s)}+1\Bigr)z_s.}}
The point\foot{Here and
in the following we denote by $z,n$ and $\rho$ the
$k$-tuples $z_1\ldots  z_k,n_1,\ldots, n_k$
and $\rho_1,\ldots,\rho_k$. We use obvious abbreviations such as
$ z^{ n}\, :=\prod_{s=1}^k z_s^{n_s}$ etc.}
$z=0$ is again a point of maximal unipotent monodromy,
and the unique powerseries solution is given
\eqn\gpowersol{
\omega_0(z)=\left. {\sum_{n_s\geq 0}c(n,\rho) z^{n+\rho}}\right|_{\rho=0},
\,\, {\rm with}\,\,
c(n,\rho)={\prod_{j}\Gamma\left(-\sum_{s=1}^k l_{0j}^{(s)} (n_s+\rho_s)+1
\right)! \over \prod_{i}\Gamma
\left(\sum_{s=1}^k l_i^{(s)} (n_s+\rho_s)+1 \right)!}\, .}
Again the system is semi resonant and the monomdromy
of \lindiffop~is reducible. Therefore one has to specify the subset
of solutions, which correspond to the $2(k+2)$
period integrals on $X^*$.
This problem was solved in \hktyII~by
factorizing the differential operators
and the following convenient basis for the period vector was found:
\eqn\pI{\Pi(z)=\left(\matrix{w_0(z)\cr
D^{(1)}_i w_0(z,\rho)|_{\rho=0}\cr
D^{(2)}_i w_0(z,\rho)|_{\rho=0}\cr
D^{(3)} w_0(z,\rho)|_{\rho=0}}\right). }
Here the $D^{(k)}_i$ are differentials with respect to the
parameter $\rho_i$, which are defined in terms of the classical
intersection numbers among the K\"ahler classes $J_i$ induced from the
$i$'th ambient space in the product space $\otimes_i {\IP}^n_i$ as follows
($\p_{\rho_i}:=\left(1\over 2 \pi i\right) \left(\p\over \p_{\rho_i}\right)$):
\eqn\defroder{
D_{i}^{(1)}:=\p_{\rho_i},\,\,\,\,
D^{(2)}_i:={\int_{X} J_i\wedge J_j\wedge J_k \over 2}\,
\p_{\rho_j}\p_{\rho_k}
\,\,\, {\rm  and} \,\,\,
D^{(3)} := -{\int_X J_i\wedge J_j\wedge J_k\over 6}  \,
\p_{\rho_i}\p_{\rho_j}\p_{\rho_k}.}
By a straightforward generalization of the monodromy requirement
one  finds the generalization of \mirrormap
\eqn\mirrormapg{t_i={\omega_i(z)\over \omega_0(z)}.}
The following explicit expansions for the correlation function
\yukawaI, which generalize \yukawaexp~
\eqn\gyuk{\langle {\cal O}_{J_i} {\cal O}_{J_j} {\cal O}_{J_k}\rangle
=\p_{t_i}\p_{t_j}{D^{(2)}_k w_0|_{\rho=0}\over w_0}(t)=
\int_X J_i \wedge J_j \wedge J_k+\sum_{d_1,\ldots,d_k} {n^r_{d_1,\ldots,d_k}
d_i\, d_j\,d_k\over
1-\prod_{i=1}^k q_i^{d_i}} \prod_{i=1}^k q_i^{d_i}}
can be read off from the period vector \pI, after normalizing by $1/w_0(z)$ and
transforming the period vector by the inverse of \mirrormapg~to the $t$
variables. The prepotential was also given in \hktyII~as
\eqn\gprepot{
F(t)=\left. { {1\over 2}\left({1\over w_0}\right)^2
\left\{ w_0 D^{(3)}w_0+D^{(1)}_l w_0 D^{(2)}_l w_0 \right\}(t)}
\right|_{\rho=0}. }
These formulas apply immediatly to all nonsingular
complete intersections in weighted projective spaces.
Let us summarize the observations, made for these series in \hktyII:
\item a.) The mirror map \mirrormapg~as well as its inverse
          have integral expansion.
\item b.) The numbers $n^r_{d_1,\ldots,d_n}$ in \gyuk~are integers.
\item c.) The constants of the quadratic polynomial in $t_i$ of
          multimoduli prepotential are $a_{ij}=0$,
          $b_i=\left(1\over 2 \pi i\right)^2 \int_X c_2 J_i \zeta(2)$ and
          $c=\left(1\over 2 \pi i\right)^3 \int_X c_3 \zeta(3)$.
\item d.) In all cases the invariants $n^r_{d_1,\ldots d_r}$
          coincide, as far as they can be checked, with the
          invariants of rational curves calculated with classical methods of
          algebraic geometry. For example, consider the
Calabi-Yau manifolds defined by
\eqn\tianyau{
p_1=\sum a_{ijk} y_i y_j y_k=0,\,\,\,p_2=\sum b_{ijk} x_i x_j x_k=0,
\,\,\, p_3=\sum c_{ij} y_i z_j=0}
as complete intersections in $\IP^3\times \IP^3$, where $y_i$ are
the homegeneous coordinates of the first $\IP^3$ and $z_i$
of the second. Then one obtains from \gyuk~the following invariants
$n^r_{d_1,d_2}$ for the rational curves of bidegree less then $6$:
$$
\vbox{\offinterlineskip
\hrule
\halign{ &\vrule# & \strut\quad\hfil#\quad\cr
\noalign{\hrule}
height1pt&\omit&   &\omit&\cr
&(0,1)&& 81 &\cr
&(0,2)&& 81 &\cr
&(0,3)&& 18 &\cr
&(0,4)&& 81 &\cr
&(0,5)&& 81 &\cr
&(0,6)&& 18 &\cr
height1pt&\omit&   &\omit&\cr
\noalign{\hrule}}
\hrule}
\quad
\vbox{\offinterlineskip
\halign{ &\vrule# & \strut\quad\hfil#\quad\cr
\noalign{\hrule}
\noalign{\hrule}
height1pt&\omit&   &\omit&\cr
&(1,1)&&  729 &\cr
&(2,2)&&  33534 &\cr
&(3,3)&& 5433399  &\cr
height1pt&\omit&   &\omit&\cr
\noalign{\hrule}
\noalign{\hrule}
height1pt&\omit&   &\omit&\cr
&(1,2)&& 2187  &\cr
&(2,4)&& 1708047 &\cr
height1pt&\omit&   &\omit&\cr
\noalign{\hrule}}
\hrule}
\quad
\vbox{\offinterlineskip
\halign{ &\vrule# & \strut\quad\hfil#\quad\cr
\noalign{\hrule}
\noalign{\hrule}
height1pt&\omit&   &\omit&\cr
&(1,3)&&  6885 &\cr
height1pt&\omit&   &\omit&\cr
\noalign{\hrule}
\noalign{\hrule}
height1pt&\omit&   &\omit&\cr
&(1,4)&&  18954&\cr
height1pt&\omit&   &\omit&\cr
\noalign{\hrule}
\noalign{\hrule}
height1pt&\omit&   &\omit&\cr
&(1,5)&&  45927 &\cr
height1pt&\omit&   &\omit&\cr
\noalign{\hrule}
\noalign{\hrule}
height1pt&\omit&   &\omit&\cr
&(2,3)&&  300348 &\cr
\noalign{\hrule}}
\hrule}
$$
\vskip 3 mm
The invariants for bidegree less then three
coincide with the ones calculated by classical methods
in \ref\sommervoll{D. E. Sommervoll, {\sl Rational Curves of low degree
on a complete intersection Calabi-Yau threefold in $P^3\times P^3$},
Oslo preprint ISBN 82-553-0838-5}.

In the remaining sections we want to investigate both the mirror
map $z(t)$ and the prepotential $F(t)$. An important question
is: are there any natural differential equations which
govern  $z$ and $F$? The answer to this questions is
affirmative as we shall see.

\newsec{Differential Equation for the Mirror Map by Examples}

We will discuss in three examples in dimensions 1,2 and 3 respectively, the
differential equation which governs the mirror map. We will state some
general properties of the equation. Our original motivation for studying
this equation was to understand the observations made experimentally on the
mirror map and the Yukawa couplings.

\subsec{Periods of Elliptic Curves}

As a warm-up, we will first consider the most elementary example of Mirror
Symmetry -- for complex curves \ref\VerlindeWarner{E. Verlinde and N. Warner,
\plb269(1991)96}
\ref\KlemmTheisenSchmidt{A. Klemm, S. Theisen and M. Schmidt ,
\ijmpa7(1992)6215}. This will be a brief exposition of some
well-known classical construction -- but in the context of Mirror
Symmetry.

Consider the following one-parameter family of
cubic curves in ${\IP}^2$:
\eqn\dumb{X_s: x_1^3+x_2^3+x_3^3-sx_1x_2x_3=0.}
We may transform $X_s$ by a $PGL(3,{\bf C})$ transformation to an
elliptic curve in the Weierstrass form:
\eqn\dumb{y^2=4x^3-g_2x-g_3}
where
\eqn\dumb{\eqalign{
g_2&=3s(8+s^3) \cr
g_3&=8+20s^3-s^6.}}
We would like to consider the variation of the
period of the holomorphic 1-form
$dx\over y$ along a homology cycle $\Gamma$:
\eqn\dumb{\omega_\Gamma=\int_\Gamma {dx\over y}.}
It can be shown that as a function of $s$, $\omega_\Gamma$ satisfies
the second order ODE:
\eqn\generalform{{d^2 \omega_\Gamma\over ds^2}+a_1(s){d\omega_\Gamma\over ds} +
a_0(s) \omega_\Gamma=0}
where
\eqn\dumb{\eqalign{
a_1&=-{d\over ds} log\left({3\over 2\Delta}(2 g_2 {dg_3\over ds} -3{dg_2\over
ds} g_3)\right)\cr
a_0&={1\over 12} a_1 {d\over ds} log\ \Delta + {1\over 12}{d^2\over ds^2}log\
\Delta - {1\over 16\Delta}(g_2 {dg_2\over ds}^2 - 12 {dg_3\over ds}^2)},}
where $\Delta=g_2^3-27 g_3^2$ is the discriminant of the above elliptic curve.
By a change of coordinate $s\rightarrow z=s^{-3}$, equation
\generalform~transforms into the hypergeometric
equation \pfN~for $N=3$ with regular singularities at $z=0,1/{3^3},\infty$:
\eqn\zform{
\left(\theta^2 - 3 z (3\theta+2)(3\theta+1)\right) \omega_\Gamma=0.}
Thus the period $\omega_\Gamma$ is a linear combination of two standard
hypergeometric functions.

We now do the following change of coordinates $s\rightarrow
J={g_2^3\over\Delta}$, and write $\omega_\Gamma$ as $\sqrt{g_2\over
g_3}\Omega_\Gamma$. Then our equation \generalform~becomes
\eqn\universal{{d^2\Omega_\Gamma \over dJ^2} +{1\over J}{d\Omega_\Gamma\over
dJ} +{31 J- 4\over 144 J^2(1-J)^2 }\Omega_\Gamma=0.}
This equation has the following universal property:
it is derived without the use of the explicit form of $g_2,g_3$ above, despite
the fact that we begun with a particular realization
(as a cubic in ${\IP}^2$) of an elliptic curve. This means that if we have
started from any other model for an elliptic curve, we will have arrived at the
same equation \universal, ie. this is the universal form of the Picard-Fuchs
equation for the periods of the elliptic curves. Note also that under the above
transformation, the ratio $t$ of two periods $\omega_\Gamma, \omega_{\Gamma'}$
(which are two hypergeometric functions) remains the same.

We can now ask for a differential equation which governs the function $t(J)$
(which is a Schwarzian triangular function). This is the well-known Schwarzian
equation: %
\eqn\elliptic{\{ t,J\} = 2\left( {3\over 16(1-J)^2} + {2\over 9J^2} + {23\over
144J(1-J)} \right).}
Here $\{z,x\}$ denotes the Schwarzian derivative ${z'''\over z'} - {3\over 2}
\left( z''\over z' \right)^2.$
Note that in this equation, by inverting $t(J)$ we may regard $J(t)$ as the
dependent variable. Recall that the inverse function for the period ratio is
precisely the mirror map. Thus $J(t)$ is our mirror map in this case and
\elliptic is our differential equation which governs it. With a suitable choice
of the period ratio $t$, $J(t)$ admits, up to overall constant,
 an integral $q$-series ($q=exp(2\pi i t)$) expansion
\eqn\jfunction{J(q)={1\over 1728}(q^{-1}+744+196884 q + 21493760 q^2 + ...).}
We can also relate the $J$-function for different realizations of the
elliptic curves in different ways to solutions of GKZ systems.
For example there exist three realizations of the elliptic curves
as hypersurfaces in weighted projective spaces $\IP^2(1,1,1),\IP^2(1,1,2)$
and $\IP^2(1,2,3)$.
$$
\vbox{\offinterlineskip\tabskip=0pt
\halign{\strut\vrule#
&\hfil~$#$
&\vrule#
&~~$#$~~\hfil
&~~$#$~~\hfil
&~~$#$~~\hfil
&\vrule#\cr
\noalign{\hrule}
& && {\rm constraint} &{\rm  diff.\,\, operator}& 1728 J(z)& \cr
\noalign{\hrule}
&P_8 && x_1^3+x_2^3+x_3^3-z^{-1/3}x_1x_2x_3=0&
\theta^2- 3 z ( 3 \theta+2)(3\theta+1)&
\displaystyle{(1+ 216z)^3\over z (1-27 z)^3} &\cr
&X_9 && x_1^4+x_2^4+x_3^2-z^{-1/4}x_1x_2x_3=0&
\theta^2- 4 z ( 4 \theta+3)(4\theta+1)&
\displaystyle{(1+ 192z)^3\over z (1-64 z)^2} &\cr
&J_{10} && x_1^6+x_2^3+x_3^2-z^{-1/6}x_1x_2x_3=0&
\theta^2- 12 z ( 6 \theta+5)(6\theta+1)&
\displaystyle{1\over z (1-432 z)} &\cr
\noalign{\hrule}}
\hrule}$$
Here the differential operators are specified by factorizing the
obvious differential operators from the general expression \lindiffop.
By the expression for the $J(z)$-function, which were obtained
by transfoming the contraints into the Weierstrass form, they can be
brought in the form \universal. The mirror map is related to the solutions
of the GKZ system, by the formulas \powsol-\mirrormap~ using
the generators of the lattice $l$ given by \cpislor. Concretely this yields,
by inversion of \mirrormap, the following expansion
for the functions $z(q)$
\eqn\mirrormapexpansions{
\eqalign{
P_8 :\,  z(q)&= q -15 q^2+ 171 q^3 -1679q^4 +
15054q^5-126981 q^6+\ldots\cr
X_9  :\,  z(q)&=q -40q^2 +1324 q^3 -39872q^4 +1136334 q^5 -31239904 q^6
      +\ldots \cr
J_{10} :\,  z(q)&= q -312q^2+ 87084 q^3 -23067968 q^4 +5930898126 q^5
      -1495818530208 q^6 +\ldots}}
The remarkable fact is that this expansions are already integer.
Inserting them into the expressions for the $J(z)$ functions
yields of course the expansion \jfunction.

The above construction (ie. the periods, the Picard-Fuchs equation
and the Schwarzian equation for the elliptic curves) is of course
classical. We will now give a similar construction for K3 surfaces
(using quartics in ${\IP}^3$) and for the quintics in ${\IP}^4$.
At the end, we will have a Schwarzian equation which governs the
period ratio (hence the Mirror map) in each of the cases. To our
knowledge, this equation is new. Actually, we also have a similar
construction for any Calabi-Yau complete intersection in a toric
variety. But for the purpose of exposition, we must restrict ourselves
to the above simple examples. Details for the general cases will be
given in our forth-coming papers\ref \KLSY{A.O. Klemm, B.H. Lian,
S.S. Roan and S.T. Yau, Differential Equations from
Mirror Symmetry I, II., in preparation.}.

\subsec{Periods of K3 surfaces}

We consider the following one-parameter family of quartic hypersurfaces in
${\IP}^3$:
\eqn\dumb{X_s: W_s(x_1,x_2,x_3,x_4)=x_1^4+x_2^4+x_3^4+x_4^4-
sx_1 x_2 x_3 x_4=0.}
The period of a holomorphic 2-form along a homology 2-cycle
$\Gamma_i$ in $X_s$ is given by \period~with $N=4$.
The Picard-Fuchs equation \pfN~for the $K_3$ case it is a third
order ODE of Fuchsian type and has singularities at $z=0,1/.{4^4},\infty$.
Thus the period $\omega_{\Gamma_i}$ is a linear combination of three
generalized hypergeometric functions. There is one solution which is
regular at $z=0$. The other two given by \logsolutions~have
singular behavior $log\ z$ and $(log\ z)^2$ respectively.

What is the analogue of the universal equation \universal~in the case
of K3 surfaces, ie. the Picard-Fuchs equation which is independent of
the model for the K3 surfaces? To answer this, we should first
interprete \universal as follows. Given a topological type of
complex n-folds $X$,
there is a universal moduli space $M$ of complex
structures on $X$. In the case of Calabi-Yau (or elliptic curves),
there is a flat Gauss-Manin connection $\nabla_M$
on $M$. The period vector ${\bf \Omega}$ of the holomorphic $n$-form
of $X$ is then a section which satisfies
\eqn\flat{\nabla_M {\bf \Omega}=0}
on a vector bundle $H^n(X,{\bf C})\rightarrow E\rightarrow M$.

In the case of the elliptic curves, we may viewed $J$ as the coordinate
on $M$. The universal Picard-Fuchs equation
\universal~should be thought of as the equation \flat in the local
coordinate. It is an interesting problem to derive the analogue of such an
equation in the case of K3 surfaces.

However in the absence of such an equation, we can still ask for the analogue
of the Schwarzian equation, ie. a differential equation for the Mirror map
which in this case is the local inverse of the function
$t(z)=\omega_1(z)/\omega_0(z)$. To write this equation, it is convenient to
first transform the Picard-Fuchs equation \pfN~to the form $({d^3\over dz^3} +
q_1(z) {d\over dz} + q_0(z))f$. This is obtained from \pfN~by a suitable change
of dependent variable
$\omega_\Gamma\rightarrow f$.
Then for the quartic model of K3 surfaces above, the Schwarzian equation is the
following fifth order ODE:
\eqn\kthree{\eqalign{
\{ z , t \}_3 =&
(-24 T_2^2+ 6 T_4)q_1{z'}^2
- 18 T_2 q_1^2{z'}^4
-4 q_1^3{z'}^6
+12 T_3 (\partial_zq_1){z'}^3\cr
&+ 3 (\partial_zq_1)^2{z'}^6-
12 T_2 (\partial^2_z q_1) {z'}^4
-6 q_1(\partial^2_z q_1)^2{z '}^6
-54 T_3q_0{z'}^3 \cr
&-27q_0^2{z'}^6
+ 36 T_2(\partial_zq_0){z'}^4
+ 18 (\partial_zq_0)q_1{z'}^6}}
where
\eqn\dumb{\eqalign{
\{ z , t \}_3 :=& -8\, T_2^3 -15\,T_3^2 + 12\, T_2\, T_4 \cr
T_i:=&\nabla^{i-2} \{z,t\}
}}
and $\nabla:=\left({d\over dt} - k {z''\over z'}\right)$. Note that prime here
means ${d\over dt}$.
For each $k$ the object $T_k dt^k$ is a rank $k$ tensor under
linear fractional transformations $t\rightarrow {a t + b\over ct + d}$, with
$a,b,c,d\in \IC$.
Then $\nabla$ above is a covariant derivative on this tensor.
The eqn \kthree has a solution given by the mirror map:
\eqn\dumb{z(q)= q - 104 q^2 + 6444 q^3 - 311744 q^4 + 13018830 q^5 - 493025760
q^6+...}

As for the classical Schwarzian equation, the new equation \kthree~is of
course $SL(2,{\bf C})$ invariant. This implies that if $z(t)$ solves the
equation, so does $z((at+b)/(ct+d))$ where $a,b,c,d$ are entries of a usual
$SL(2,\IC)$ matrix. Beside the invariance under this linear fractional
transformation
the differential equation \kthree~exhibits also invariances under
nonlinear transformations, which were used in \KLSY~to fix the
numerical coefficients
in \kthree\dumb~uniquely. Once again we have observed experimentally that the
$q$-series expansion of the mirror map $z$ which satisfies \kthree is in fact
integral.
In the case of elliptic curves (using the Weierstrass model), we have seen that
the mirror map is given by the $J$ function which is well-known to have an
integral expansion.
It would be interesting to establish a similar statement for $z(q)$
in the case of K3.

\subsec{Periods of Quintic Threefolds}

The periods of the quintic hypersurface in ${\IP}^4$ were studied
in the last section. Special geometry introduces the prepotential
$F$ as the new object of interest. The Weil-Peterson metric on the complex
structure moduli space of mirror of the quintic $X^*$ is described by $F$.
Moreover
the mirror hypothesis asserts that there is a special coordinate
transformation given by a ratio of periods $t=\omega_1(z)/\omega_0(z)$, in
which $\p_t^3 F(t)$ gives the generating function for the number of rational
curves in a generic quintic.
It is therefore important to understand both the mirror map $z$ and the
prepotential $F$. Thus a relevant
question is: are there natural differential equations
which govern $z$ and $F$?

For the mirror map $z$, there is a natural generalization of the Schwarzian
equations \elliptic \kthree. Specifically, we claim that the mirror map $z(t)$
defined above satisfies the following seventh order ODE (see next section):
\eqn\explicitschwarzIV{\{z,t\}_4=-256 q_0^3z'^{12}+ 128 q_0^2q_2^2z'^{12} +
...}
where
\eqn\scharzIV{\eqalign{
\{ z , t \}_4  := & -64T_2^6 -560T_2^3T_3^2 -1275T_3^4 + 448T_2^4T_4
              +2040 T_2T_3^2T_4 - 192 T_2^2T_4^4 - \cr
             & 504 T_4^3 + 1120T_2^2T_3T_5
              + 840T_3T_4T_5 - 280 T_2T_5^2 + 20\,T_6\,\{z,t\}_3.}}
As in the cases of K3 surfaces and elliptic curves, this Schwarzian equation is
also manifestly $SL(2,\IC)$ invariant. In the case of the quintic hypersurface,
the mirror map which satisfies this equation has the
$q$-expansion:
\eqn\dumb{z(q)= q - 770 q^2 + 171525 q^3 - 81623000 q^4 - 35423171250 q^5 -
   54572818340154 q^6+...}

For the prepotential $F$, we have also derived a similar (seventh order) but
considerably more complicated polynomial differential equation.  We will
discuss this in the last section.

\newsec{Construction of the Schwarzian equations}

We now give an exposition for the construction of the differential
equation which governs our mirror map $z(t)$ in each case.

Note that in each case we begin with an $n^{th}$ order ODE of Fuchsian type:
\eqn\ode{Lf:=\left({d^n\over dz^n}+\sum_{i=0}^{n-1} q_i(z) {d^i\over
dz^i}\right)f=0}
($n$ being 2,3 and 4 respectively for the elliptic curves, K3 surfaces
and Calabi-Yau 3-folds.) In particular,
the $q_i(z)$ are rational functions of $z$.
Let $f_1, f_2$ be two linearly independent solutions of this equation
and consider the ratio $t:=f_2(z)/f_1(z)$. Inverting this relation
(at least locally), we obtain $z$ as a function of $t$. Our goal is
to derive an ODE, in a canonical way, for $z(t)$.

We first perform a change of coordinates $z\rightarrow t$ on \ode~ and
obtain:
\eqn\odeb{\sum_{i=0}^n b_i(t) {d^i\over dt^i} f(z(t))=0}
where the $b_i(t)$ are rational expressions of the derivatives $z^{(k)}$
(including $z(t)$). For example we have $b_n(t)=a_n(z(t)) z'(t)^{-n}$.
It is convenient to simplify the equation
by writing (gauge transformation)
$f=A g$, where $A=exp(-\int {b_{n-1}(t)\over n b_n(t)})$, and
multiplying \odeb~ by ${1\over A b_n}$ so that it
becomes
\eqn\odec{\tilde{L}g:=\left({d^n\over dt^n} +\sum_{i=0}^{n-2} c_i(t) {d^i\over
dt^i}\right) g(z(t))=0}
where $c_i$ is now a rational expression
of $z(t),z'(t),..,z^{(n-i+1)}$ for
$i=0,..,n-2$. Now $g_1:=f_1/A$ and $g_2:=f_2/A=t g_1$
are both solution to the equation \odec. In particular we have
\eqn\PQ{\eqalign{
P:=&\tilde{L}g_1=\left({d^n\over dt^n}+\sum_{i=0}^{n-2} c_i(t) {d^i\over
dt^i}\right) g_1=0\cr
Q:=&\tilde{L}(tg_1) -t\tilde{L}g_1=\left(n{d^{n-1}\over
dt^{n-1}}+\sum_{i=0}^{n-3} (i+1) c_{i+1}(t) {d^i\over dt^i}\right) g_1=0.}}
 Note that since $c_i$ is a rational expression of $z(t),z'(t),..,z^{(n-i+1)}$,
it follows that $P$ involves $z(t),..,z^{(n+1)}$
while $Q$ involves only $z(t),..,z^{(n)}$.
Eqns \PQ~ may be viewed as a coupled system of differential equations for
$g_1(t),z(t)$. Our goal
is to eliminate $g_1(t)$ so that we obtain an equation for just $z(t)$.
One way to construct this is as follows. By \PQ, we have
\eqn\system{\eqalign{
{d^i\over dt^i} P=&0,\ \ \ \ i=0,1,..,n-2,\cr
{d^j\over dt^j} Q=&0,\ \ \ \ j=0,1,..,n-1.}}
We now view \system~ as a homogeneous
{\it linear} system of equations:
\eqn\matrixform{\sum_{l=0}^{2n-2} M_{kl}(z(t),..,z^{(2n-1)}(t)){d^l\over
dt^l}g_1=0,\ \ \ k=0,..,2n-2,}
where each $(M_{kl})$ is the following $(2n-1)\times(2n-1)$ matrix:
\eqn\M{\left(\matrix{
c_0&c_1&..&c_{n-2}&0&1&0&..&0\cr
c_0'&c_0+c_1'&..&c_{n-3}+c_{n-2}'&.&0&1&0&0\cr
 &\cdots&..&\cdots& & & &..&\cr
c_0^{(n-2)}&(n-2)c_0^{(n-3)}+c_1^{(n-2)}&
..&\cdots&.&.&.&0&1\cr
c_1&2c_2&..&(n-2)c_{n-2}&n&0&0&..&0\cr
c_1'&c_1+2c_2'&..&(n-3)c_{n-3}+(n-2)c_{n-2}'&0&n&0&..&0\cr
 &\cdots&..&\cdots& & &..&\cr
c_1^{(n-1)}&(n-1)c_1^{(n-2)}+2c_2^{(n-1)}&..&\cdots&.&
&.&0&n}\right)}
More precisely if we define the $1^{st}$ and $n^{th}$ ($n$ fixed)
row vectors to be
$(M_{1l})=(c_0,c_1,..,c_{n-2},0,1,0,..,0)$ and
$(M_{nl})=(c_1,2c_2,..,(n-2)c_{n-2},0,n,0,..,0)$ respectively, then the matrix
$(M_{kl})$ is given by the recursion relation:
\eqn\recursion{M_{k+1,l}=M_{k,l-1}+M_{k,l}',\ \ \ \
l=1,..,2n-1;\ \ k=1,..,n-2,n,..,2n-2.}
Thus the $(M_{kl})$ depends rationally on $z(t),..,z^{(2n-1)}(t)$. Since $g_1$
is nonzero, it follows that
\eqn\det{det\left( M_{kl}(z(t),..,z^{(2n-1)}(t))\right)=0.}
This is what we call the Schwarzian equation associated with \ode. Note that
by suitably clearing denominators, this becomes a $(2n-1)^{st}$ order
polynomial ODE for $z(t)$ with constant coefficients.
It is clear that this equation depends only on the data $q_i(z)$ we began
with. In the case in which all the $q_i$ are identically zero, we call the
determinant in \det~ the $n^{th}$ Schwarzian bracket $\{z(t),t\}_n$.

Despite having a general form of the Schwarzian equation, it is useful to see a
few simple examples.
As the  first example, consider the case $n=2$:
\eqn\dumb{{d^2 \over dz^2} f + q_0(z) f=0.}
The eqns \PQ~ become
\eqn\dumb{\eqalign{
{g_1}'' + c_0 g_1 =&0\cr
2{g_1}'=&0}}
where
\eqn\dumb{c_0(t):=q_0(z(t)) {z'}^2 - {1\over 2}\{z,t\}_2.}
The corresponding linear system \matrixform~ has:
\eqn\dumb{(M_{kl})=\left(\matrix{c_0&0&1\cr
0 & 2 & 0\cr
0 & 0 & 2}\right). }
Hence the associated Schwarzian equation \det~ in this case is
\eqn\dumb{det(M_{kl})=4c_0=2\left(2q_0 {z'}^2 - \{z,t\}_2\right)=0}
which is the well-known classical Schwarzian equation.

For $n=3$, we begin with the data
\eqn\dumb{\left( {d^3\over dz^3} + q_1(z){d\over dz} + q_0(z)\right)f=0.}
The transformed equation \odec~ in this case becomes
\eqn\dumb{\left( {d^3\over dt^3} + c_1(t){d\over dt} + c_0(t)\right)g=0}
where
\eqn\ccexpression{\eqalign{
c_1(t) :=& q_1(z(t))\,{{z'(t)}^2} + 2\{z(t),t\}_2 \cr
c_0(t) :=& q_0(z(t))\,{{z'(t)}^3} + q_1(z(t))\,z'(t)\,z''(t) +
   {{3\,{{z''(t)}^3}}\over {{{z'(t)}^3}}} \cr &-
   {{4\,z''(t)\,z^{(3)}(t)}\over {{{z'(t)}^2}}} + {{z^{(4)}(t)}\over {z'(t)}}.
}}
The corresponding linear system \matrixform~ has:
\eqn\dumb{(M_{kl}) = \left(\matrix{ c_0 & c_1 & 0 & 1 & 0\cr
c_0'&c_0+c_1'&c_1&0&1\cr
c_1&0&3&0&0\cr
c_1'& c_1 & 0 & 3 & 0\cr
c_1''&2c_1'&c_1&0&3} \right).}
Computing the associated Schwarzian equation, we get
\eqn\schwarzIII{
det(M_{kl}):=27 c_0^2 + 4c_1^3-18c_1 c_0'-3{c_1'}^2+ 6c_1 c_1''=0.}
Substituting \ccexpression~ into \schwarzIII, we get the explicit form
\kthree.

Now let's consider the case $n=4$ which begins with
\eqn\dumb{\left( {d^4\over dz^4} + q_2(z){d^2\over dz^2} + q_1(z){d\over dz} +
q_0(z)\right)f=0.}
We assume that this is the Picard-Fuchs equation for the periods of a 3
dimensional Calabi-Yau hypersurface. Then as pointed out earlier, there is a
basis of solutions which takes the form \periods. This implies that
$q_1(z)={dq_2\over dz}$. The analogue of transformed equation \odec~ now
becomes
\eqn\transform{\left( {d^4\over dt^4} + c_2(t){d^2\over dt^2} + c_2'(t){d\over
dt} + c_0(t)\right)g=0}
where
\eqn\ccrr{\eqalign{
c_2(t):=& q_2(z){z'}^2 + 5\{z(t),t\}_2 \cr
c_0(t):=& q_0(z){z'}^4 +
   {3\over 2} {dq_2(z)\over dz}{z'}^2 z'' -
   {3\over 4} q_2(z) {z''}^2 -
   {{135{z''}^4}\over {16{z'}^4}} \cr+
  & {3\over 2} q_2(z) z' z^{(3)}+
   {{75{z''}^2 z^{(3)}}\over {4{z'}^3}} -
   {{15{z^{(3)}}^2}\over {4{z'}^2}} -
   {{15z'' z^{(4)}}\over {2{z'}^2}} +
   {{3z^{(5)}}\over {2z'}} }.}
The associated linear system \matrixform~ in this case is $7\times 7$.
Computing the associated Schwarzian equation \det, we get
\eqn\schwarzIV{\eqalign{
det(M_{kl}):=&16\,{{c_2 }^4}\,c_0  - 128\,{{c_2 }^2}\,{{c_0 }^2} +
   256\,{{c_0 }^3} + 4\,{{c_2 }^3}\,{{c_2 '}^2} \cr+
&   240\,c_2 \,c_0 \,{{c_2 '}^2} - 15\,{{c_2 '}^4} -
   144\,{{c_2 }^2}\,c_2 '\,c_0 ' -
   448\,c_0 \,c_2 '\,c_0 ' \cr+
&   256\,c_2 \,{{c_0 '}^2} - 8\,{{c_2 }^4}\,c_2 '' +
   128\,{{c_0 }^2}\,c_2 '' -
   48\,c_2 \,{{c_2 '}^2}\,c_2 '' \cr+
&   48\,c_2 '\,c_0 '\,c_2 '' +
   12\,{{c_2 }^2}\,{{c_2 ''}^2} -
   48\,c_0 \,{{c_2 ''}^2} + 32\,{{c_2 }^3}\,c_0 '' \cr-
&   128\,c_2 \,c_0 \,c_0 '' +
   48\,{{c_2 '}^2}\,c_0 '' +
   32\,{{c_2 }^2}\,c_2 '\,c_2 ^{(3)} +
   64\,c_0 \,c_2 '\,c_2 ^{(3)} \cr-
&   96\,c_2 \,c_0 '\,c_2 ^{(3)} +
   8\,c_2 \,{{c_2 ^{(3)}}^2} -
   8\,{{c_2 }^3}\,c_2 ^{(4)} +
   32\,c_2 \,c_0 \,c_2 ^{(4)} -
   12\,{{c_2 '}^2}\,c_2 ^{(4)}=0.\cr
}}
Substituting \ccrr~ into \schwarzIV~, we get the 7th order ODE
\explicitschwarzIV.

\newsec{ODE for the Instanton corrected Yukawa Coupling}

We can give an analogous construction of an ODE for the Yukawa coupling
\yukawaI.
To be brief, we will instead explain an approach which
will in the end results in a simple {\it characterization}.

Let's consider first a classical problem: given a pair of polynomials
$r(y,z),s(y,z)$, let $X$ be the intersection of their zero loci (in ${\bf
C}^2$) and $X_y,X_z$ be the projections of $X$ onto the $y,z$-directions. When
can we construct (by quadrature) nontrivial polynomials $p(y),q(z)$ whose zero
loci contain $X_y,X_z$ respectively? The answer is simple: Hilbert's
Nullstellensatz gives the following characterization. Namely $p(y),q(z)$ are
constructible iff $X$ is finite. Moreover there exists canonical choices for
$p(y),q(z)$, namely the ones with the minimal
degrees. Unfortunately, no similar general characterization is known in the
case of {\it differential} polynomials (see however \ref\ritts{J.F. Ritt,
Differential Algebra, Colloq. Publ. vol.33, AMS, Providence, R.I., 1950.}).
However in the cases we consider, we can formulate our problem of constructing
an ODE for the Yukawa coupling in a similar spirit. We can also construct the
analogues of the $p(y),q(z)$ above.

We begin with \transform and \ccrr. By applying \transform to the four
solutions of the form $u(t), u(t)t, u(t)F'(t), u(t)(2F(t)-tF'(t))$ (cf. eqn
\periods), we get a system of four equations which is equivalent to:
\eqn\uKeqn{\eqalign{u^{(4)}+c_2 u''+c'_2 u' + c_0 u=&0\cr
4u'''+2c_2 u'+ c_2' u=&0\cr
F^{(3)}(6u'' + c_2 u) +4F^{(4)}u'+F^{(5)}u=&0\cr
2F^{(3)}u'+F^{(4)}u=&0.\cr }}
Solving the last equation gives $u=(F^{(3)})^{-1/2}$. The second equation is
redundent because it is the derivative of the third one. Thus the above system
reduces to just
\eqn\zKeqn{\eqalign{
&A_2(z;t)-B_2(y;t)=0\cr
&A_4(z;t)-B_4(y;t)=0\cr
}}
where $A_2(z;t),A_4(z;t),B_2(y;t),B_4(y;t)$ are differential rational functions
defined by
\eqn\aabb{\eqalign{
A_2(z;t):=&c_2(t)= q_2(z){z'}^2 + 5\{z(t),t\}_2 \cr
A_4(z;t):=&c_0(t)= q_0(z){z'}^4 +
   {3\over 2} {dq_2(z)\over dz}{z'}^2 z'' -
   {3\over 4} q_2(z) {z''}^2 -
   {{135{z''}^4}\over {16{z'}^4}} \cr+
  & {3\over 2} q_2(z) z' z^{(3)}+
   {{75{z''}^2 z^{(3)}}\over {4{z'}^3}} -
   {{15{z^{(3)}}^2}\over {4{z'}^2}} -
   {{15z'' z^{(4)}}\over {2{z'}^2}} +
   {{3z^{(5)}}\over {2z'}} \cr
B_2(y;t) :=& 2 y'' - {{y'}^2 \over 2}\cr
B_4(y;t):=& {y^{(4)}\over 2} + {{y''}^2\over 4} -
{y'' {y'}^2\over 2}+{{y'}^4\over 16}\cr
y(t):=& Log\ F'''(t).
}}
The system \zKeqn~ depends only on the two rational functions $q_2(z),q_0(z)$
via $A_2,A_4$, whereas $B_2,B_4$ are independent of such data.
Observe that if we assign a weight 0 to $z,y$ and weight 1 to $d\over dt$, then
the expressions $A_2,B_2$ (resp. $A_4,B_4$) are formally homogeneous of weights
2 (resp. 4). Thus we can speak of the weight of a quasi-homogeneous
differential polynomial of these expressions.
Finally, we note that the system of equations \zKeqn~ should be regarded as
a differential analogue of the system of polynomial equations $r(y,z)=0=s(y,z)$
considered above. With this analogy in mind, we now state the first of the two
main results of this section:

\vskip .3in
{\it Given a pair of rational function $(q_0(z),q_2(z))$ (which determines the
Picard-Fuchs equation), there exists a differential polynomial $P$ with the
following properties:

(i) $P$ quasi-homogeneous;

(ii) $P(A_2(z;t),A_4(z;t))$ is identically zero (see eqns. \aabb);

(iii) $P(B_2(y;t),B_4(y;t))=0$ is a nontrivial 7th order ODE in $y$ which has a
solution $y(t)=Log\ F'''(t)$ where $F(t)$ is the prepotential;

(iv) $P$ is minimal, ie. every differential polynomial satisfying (i)--(iii)
has weight no less than that of $P$.

(v) The polynomial ODE $P(B_2(y;t),B_4(y;t))=0$ is $SL(2,{\bf C})$ invariant.}
\vskip .3in

We have observed in all the known examples that $P$ is in fact characterized by
the above properties, ie. $P$ satisfying (i)-(v) is unique up to constant
multiple. We conjecture that this is the case in general.
Note that $P$ depends on the data $(q_0(z),q_2(z))$ precisely via property
(ii).
The polynomial ODE $P(B_2(y;t),B_4(y;t))=0$ should be regarded as the
differential analogue of $p(y)=0$ considered above. In this analogy, the group
$SL(2,{\bf C})$ plays the role for the {\it differential} polynomial
$P(B_2(y;t),B_4(y;t))$ as the Galois group does for the {\it ordinary}
polynomial $p(y)$.
The simplest example of the above ODE for $F(t)$ is given by the considering
the Picard-Fuchs equation for the complete intersection of 4 quadrics in ${\bf
P}^7$, which is given by \lindiffop~after factorizing
$\theta^4$ as $\left[\theta^4-16 ( 2\theta-1)^4\right] \omega_\Gamma=0$.
To write the ODE for $F(t)$ down, we first define the notations:
\eqn\dumb{\eqalign{
\rho=&100B_4- 9B_2^2- 30B_2''\cr
\chi=&-32\rho^2 B_2 -45{\rho'}^2 +40\rho\rho''\cr
\delta=&5\chi\rho'- 2\chi'\rho.}}
Then our ODE for the prepotential in this case is
\eqn\dumb{
{\ninepoint{\eqalign{
&  3783403212890625\,{{\chi }^{18}} +
   52967644980468750\,{{\chi }^{15}}\,{{\delta }^2} +
   292835408677734375\,{{\chi }^{12}}\,{{\delta }^4}\cr & +
   833559395864062500\,{{\chi }^9}\,{{\delta }^6} +
   1301823644717109375\,{{\chi }^6}\,{{\delta }^8} +
   1064406315612768750\,{{\chi }^3}\,{{\delta }^{10}} \cr & +
   357449882108765625\,{{\delta }^{12}} +
   9097175898878906250\,{{\chi }^{16}}\,{{\rho }^5} +
   75543680906950781250\,{{\chi }^{13}}\,{{\delta }^2}\,{{\rho }^5} \cr & -
   55168781762820937500\,{{\chi }^{10}}\,{{\delta }^4}\,{{\rho }^5}-
   1235933279927738437500\,{{\chi }^7}\,{{\delta }^6}\,{{\rho }^5} \cr &-
   2628328829388247068750\,{{\chi }^4}\,{{\delta }^8}\,{{\rho }^5} -
   1669442421173622243750\,\chi \,{{\delta }^{10}}\,{{\rho }^5} \cr &-
   316395922222462973709375\,{{\chi }^{14}}\,{{\rho }^{10}} -
   1041303693581386404075000\,{{\chi }^{11}}\,{{\delta }^2}\,{{\rho }^{10}}
   \cr &+
   1397061241390545045311250\,{{\chi }^8}\,{{\delta }^4}\,{{\rho }^{10}} -
   978071752628929206000\,{{\chi }^5}\,{{\delta }^6}\,{{\rho }^{10}} \cr & +
   3088961515882945520173125\,{{\chi }^2}\,{{\delta }^8}\,{{\rho }^{10}} +
   726375263921582813504122500\,{{\chi }^{12}}\,{{\rho }^{15}} \cr &-
   2401967567306257982918892000\,{{\chi }^9}\,{{\delta }^2}\,{{\rho }^{15}} +
   2931906039367569842399977800\,{{\chi }^6}\,{{\delta }^4}\,{{\rho }^{15}} \cr
& -
   2592007729730548310729752000\,{{\chi }^3}\,{{\delta }^6}\,{{\rho }^{15}}+
   26477211431856325292132500\,{{\delta }^8}\,{{\rho }^{15}} \cr & +
   731773527868504699561324929024\,{{\chi }^{10}}\,{{\rho }^{20}} -
   1384453886791545382987331665920\,{{\chi }^7}\,{{\delta }^2}\,
    {{\rho }^{20}} \cr & + 1003786188392583028918031769600\,{{\chi }^4}\,
    {{\delta }^4}\,{{\rho }^{20}} \cr & -
   92650299984331138894225408000\,\chi \,{{\delta }^6}\,{{\rho }^{20}} \cr &+
   264379950716374035480557566033920{{\chi }^8}\,{{\rho }^{25}}\cr & -
   323653884996678359415539902709760{{\chi }^5}\,{{\delta }^2}
    {{\rho }^{25}}  \cr
  &+105122101152057682020817226956800\,{{\chi }^2}\,
    {{\delta }^4}\,{{\rho }^{25}} \cr & +
   48853700167414249640038923438653440\,{{\chi }^6}\,{{\rho }^{30}} \cr & -
   33153423760664989683513831467253760\,{{\chi }^3}\,{{\delta }^2}\,
    {{\rho }^{30}}\cr & + 528120679253988321156369324441600\,{{\delta }^4}\,
    {{\rho }^{30}} \cr & + 4965538896010513223822010617996247040\,{{\chi }^4}\,
    {{\rho }^{35}} \cr &- 1238934080748073699029086124292177920\,\chi \,
    {{\delta }^2}\,{{\rho }^{35}}\cr & +
   265021771162266355900761945816768184320\,{{\chi }^2}\,{{\rho }^{40}}
   \cr & +
   5822406825670998196401392296588763725824\,{{\rho }^{45}}=0
}}}}

Note that since $\delta,\chi,\rho$ are of weights 15,10,4 respectively,
$P$ is a quasi-homogeneous differential polynomial of weight 180.
Each of the 37 terms in this
polynomial corresponds to a partition of 180 by 15,10,4.

It turns out that there is a dual characterization for the Schwarzian equation
\schwarzIV we have constructed:
\vskip .3in
{\it There exists a differential polynomial $Q$ with the following properties:

(i) $Q$ is quasi-homogeneous;

(ii) $Q(B_2(y;t),B_4(y;t))$ is identically zero (see eqns. \aabb);

(iii) $Q(A_2(z;t),A_4(z;t))=0$ is a nontrivial ODE which has a solution given
by the mirror map $z(t)$;

(iv) $Q$ is minimal of weight 12, ie. every differential polynomial satisfying
(i)--(iii) has weight at least 12;

(v) The polynomial ODE $Q(A_2(z;t),A_4(z;t))=0$ is $SL(2,{\bf C})$ invariant;

(vi) $Q$ is universal, ie. it is independent of the data $(q_2(z),q_0(z))$ and
it is characterized by (i)-(v) up to constant multiple;

(vii) $Q(A_2(z;t),A_4(z;t))=0$ coincides with $det(M_{kl})=0$ in \schwarzIV. }
\vskip .3in
We will defer the detailed proofs of the above results to our forth-coming
papers.

\vskip.5cm
\noindent
{\bf Acknowledgement:} Two of us (A. K. and S-T. Y.) would like to
thank Shinobu Hosono and Stefan Theisen for collaboration on the
matters discussed in section 2. B.H.L. thanks Gregg Zuckerman for
helpful discussions.

\listrefs
%\listfigs   %(if necessary)
\bye